\documentclass[traditabstract]{aa} % traditional abstract                                   
\usepackage{longtable}
\usepackage{graphicx}
\usepackage{txfonts}
 
\begin{document}
   \title{AGILE Mini-Calorimeter gamma-ray burst catalog}

   \author{M. Galli
          \inst{1}\and
          M.Marisaldi
          \inst{2}\and
          F.Fuschino
          \inst{2}\and 
          C.Labanti
          \inst{2}\and 
          A.Argan
          \inst{4}\and 
          G.Barbiellini
          \inst{8,13}\and  
          A.Bulgarelli
          \inst{2}\and                              
          P.W.Cattaneo
          \inst{7}\and  
          S.Colafrancesco
          \inst{15,16}\and                    
          E.Del Monte          
          \inst{3}\and 
          M.Feroci          
          \inst{3}\and 
          F. Gianotti         
          \inst{2}\and                     
          A. Giuliani         
          \inst{12}\and           
          F. Longo
          \inst{8}\and 
          S.Mereghetti         
          \inst{12}\and                   
          A. Morselli          
          \inst{5}\and 
          L. Pacciani       
          \inst{3}\and           
          A. Pellizzoni         
          \inst{6}\and 
          C. Pittori         
          \inst{15,9}\and  
          M. Rapisarda     
          \inst{11}\and                              
          A. Rappoldi        
          \inst{7}\and 
          M.Tavani
          \inst{3,14}\and                      
          M.Trifoglio
          \inst{2}\and  
          A.Trois 
          \inst{6} \and 
          S. Vercellone 
          \inst{10} \and           
          F. Verrecchia  
          \inst{15,9}              
          }

   \institute{ENEA Bologna, via Martiri di Montesole 4, I-40129 Bologna, Italy \\  % 1 
              \email{marcello.galli@enea.it}
         \and
             INAF/IASF Bologna, via Gobetti 101, I-40129 Bologna, Italy            % 2 
         \and    
             INAF/IAPS, via Fosso del Cavaliere 100, I-00133 Roma, Italy           % 3 
         \and 
             INAF, viale del Parco Mellini 84, I-00136 Roma, Italy                 % 4 
         \and 
             INFN, Roma Tor Vergata, via della Ricerca Scientifica 1, I-00133 Roma, Italy  % 5 
         \and  
             INAF, Osservatorio Astronomico di Cagliari, localit\`a Poggio dei Pini, strada 54, I-09012 Capoterra, Italy % 6         
         \and
             INFN Pavia, via Bassi 6, I-27100 Pavia, Italy        % 7 
         \and   
             INFN Trieste, Padriciano 99, I-34012 Trieste,Italy  % 8 Franz Longo  
         \and  
             ASI Science Data Center, via G. Galilei, I-00044 Frascati, Italy   % 9 
         \and  
             INAF/IASF Palermo, Via Ugo La Malfa 153, I-90146 Palermo, Italy % 10    
         \and 
             ENEA Frascati,  via Enrico Fermi 45, I-00044 Frascati, Italy    % 11 Rapisarda              
         \and
             INAF/IASF Milano, via E. Bassini 15, I-20133 Milano, Italy  %12 
         \and
             Dip. di Fisica, Universit\`a di Trieste, via Valerio 2, I-34127 Trieste, Italy  % 13 
         \and   
             Dip. di Fisica, Universit\`a ``Tor Vergata'', via della Ricerca Scientifica 1, I-00133 Roma, Italy  % 14 
         \and   
             INAF, Osservatorio Astronomico di Roma, Via Frascati 33, I-00040 Monteporzio, Italy  %15
         \and                                  
             School of Physics, University of the Witwatersrand, Private Bag 3, WITS-2050 Johannesburg, South Africa  %16
             }

   \date{Received 3 December 2012 / Accepted 19 February 2013 }
 
  \abstract
  { 
  The Mini-Calorimeter of the AGILE satellite 
  can observe the high-energy part of gamma-ray bursts 
  with good timing capability.
  We present the data of the 85 hard gamma-ray bursts 
  observed by the Mini-Calorimeter since the launch  (April 2007) until October 2009.
  We report the timing data for 84 and spectral data for 21 bursts.
  The data table are available from the Strasbourg Astronomical Data Center (CDS)  
  \thanks{Tables 1, 2, and 4 are available in electronic form
   via anonymous ftp to cdsarc.u-strasbg.fr (130.79.128.5)
   or via http://cdsweb.u-strasbg.fr/cgi-bin/qcat?J/A+A/}
  and the detailed data from the ASI Science Data Center (ASDC)
  \thanks{Web site: http://agile.asdc.asi.it/}.
  \vspace{0.5cm plus 0cm minus 0cm}
  }
  
   \keywords{gamma rays:observations --   
                   X-rays: bursts -- 
                   Catalogs
              }   
   \maketitle
%
%________________________________________________________________

\section{Introduction} 

In recent years many instruments have been devoted to the study of
the gamma-ray bursts (GRBs), which remain one of the most puzzling 
phenomena in the Universe. 

The energy released from these events spans
a wide spectral region, from radio to GeV, 
with great variations in temporal and spectral behavior (\cite{mezaros}). 
In most cases, the emission peaks in the range 50--500 keV;
for this reason most of the instruments dedicated to
GRBs are optimized for detection in this energy range.

The Mini-Calorimeter (MCAL) (\cite{mcal})
onboard the AGILE satellite (\cite{agile}) 
is an instrument able to detect gamma-rays 
from $\sim$350 keV to 100 MeV region with a time
resolution better than 2 ${\mu}$s.                         
The main task of MCAL is to support the AGILE silicon tracker (\cite{stp} , \cite{stb})
in measuring the gamma-ray energy, but it is also used
as an independent transients monitor and to investigate the
high-energy part of GRBs (\cite{mcalgrb}).

Onboard the AGILE satellite the SuperAGILE (SA) instrument (\cite{sa}) 
also acts as a GRB monitor. It is a coded
mask telescope operating in the energy range 15--45 keV  with
an angular resolution of 6 arcmin.  
Because of the very different energy ranges and 
the larger MCAL field of view (FOV)
only a small fraction of the GRBs are detected by both instruments. 
The GRB data collected by SA are not homogeneous 
with the MCAL sample and are not considered in this paper.

Here we present a homogeneous sample of  GRB data collected 
by MCAL between February 2008 and October 2009.
During this period AGILE was operating in {\em pointing mode}, therefore
a unique incoming direction in the reference frame of the spacecraft
can be associated to each GRB. 
Since October 2009, AGILE is operating in {\em spinning mode},
due to a reaction-wheel failure,
scanning a large fraction of the sky with a typical
angular velocity of  a few degrees per second. 
The GRB data collected in spinning mode 
need different analysis methods and will be the 
subject of a forthcoming study.

In Section 2 we briefly describe the MCAL instrument,
in Section 3 we describe the data collection and processing,
Section 4 deals with the temporal analysis of the GRBs,
Section 5 deals with the spectral analysis, and Section 6 
presents our conclusions. 

%__________________________________________________________________

\section{Instrument}

The Mini-Calorimeter of the AGILE Satellite has been fully described 
elsewhere (\cite{mcal}): it consists of an array of 30 CsI(Tl) bars, 
each one of size $ 15 \times 23 \times 375 \,\mathrm{mm}^3$,
arranged in two layers with the bars aligned along orthogonal directions 
for a total thickness of 3 cm.

The scintillation light produced by the interacting gamma-rays 
is collected at each end of the bars by
photo-diodes (PD); the attenuation of the light 
transmitted along the bar can be represented by 
an exponential decay; the decay coefficient has an
average value of $ 0.028\,\mathrm{cm}^{-1}$ and has been measured 
for each bar before the final assembly of the instrument.
The energy released in the bar and the event position along the bar
are obtained according to the exponential decay law, 
as described in a previous paper (\cite{mcal}).
Measurements have shown that this is a reasonable
approximation and fails only for energy released 
at the very end of the bars, where
geometrical effects dominate over attenuation.

A discriminator circuit with programmable threshold 
receives the sum of the electric signals from the PDs; 
when the sum is higher than the threshold, the PDs signals are 
sent to an ADC, time-tagged and stored in a circular buffer.
Because of the light attenuation, this electric signal threshold
does not correspond to a fixed energy threshold: 
the bars are more sensitive to low-energy events
near the photo-diodes than at the center of the bars. 
The PDs are characterized by their gain ($\mathrm{e}^-/\mathrm{keV}$)
and the zero-level signal (offset). 
These parameters have been measured for each PD before launch  and
are used to evaluate the photon energy (\cite{mcal}). 

The MCAL instrument is situated below the silicon tracker,
but its FOV is not limited to the FOV 
of the tracker and acts as an all-sky monitor.
The effective area of the detector is between 200 and 500 $\mathrm{cm}^2$,
depending on photon energy and angle (\cite {mcal}).

A complex, fully programmable trigger logic for transient event detection 
is implemented on-board (\cite{DR}).
This trigger logic is based on the counts given by a number of 
detector rate-meters (DR): four independent rate-meters are obtained
by dividing each MCAL layer into two parts and summing the
signals from the bars of each part. 
The DR counts are integrated over three energy ranges ($ < 1.4 \mathrm{MeV} , 
1.4-3.0 \mathrm{MeV}, > 3.0 \mathrm{MeV}$)
and six time windows (1, 16, 64, 256, 1024 and 8192 ms). 
These integrated counts are compared to the integrated background signal 
with a background integration-time chosen among seven time intervals
(8, 16, 32, 65, 131, 262 and 524 s) 
and delayed with respect to the DR integration times.
For the 1-ms and 16-ms time windows, the MCAL is considered as a whole
and not divided into four parts. An additional moving 
time window of $300 \, \mu $s duration is
implemented for the detection of very short transients. 

A partial trigger signal is issued for each DR if the counts
exceed a given threshold. The trigger logic compares all
partial triggers; a GRB trigger signal is issued 
only if a proper combination of partial triggers is obtained, 
otherwise the event is rejected.
Different combinations of DRs, thresholds, time windows,  
and background integration-times, can be chosen; moreover
the trigger criterion can be either static or dynamic:
in the static criterion the threshold is fixed, while 
in the dynamic criterion the  threshold is proportional to the 
standard deviation of the background. 
The GRB end time is determined by the time at which all 
DRs decrease below given thresholds; 
static and dynamic criteria are also implemented
for the GRB ending time choice (TSTOP time).
The trigger logic has been tuned during the 
first months of the mission, to optimize sensitivity and
telemetry load, and minimizing the rate of false-positive triggers.
The configuration implemented onboard requires that at least two partial 
triggers on different DR are issued at the same time for time windows 
larger than 16 ms. For these time windows the dynamic criterion was set, 
requiring that a partial trigger is issued only when the count rate 
exceeds the measured background rate by at least five standard deviations. 
The requirement for two simultaneous partial triggers implies a rather 
uniform involvement of all detection planes and helps to reject local 
enhancements generated by, for example, noise increase. For time windows of 
16 ms and shorter, the static trigger criterion was set, requiring at least 
seven, ten and ten counts in the $300\;\mu s$, 1 ms and 16 ms time windows, 
respectively.
 
When a GRB trigger is issued, data are transmitted to the 
ground station on a photon-by-photon basis.  
Together with the GRB data, many seconds of data before and after the
GRB are sent to ensure a proper estimate of the
background level.

Since the end of the commissioning phase, MCAL and
  SuperAGILE, are part of the third InterPlanetary
  Network (IPN)\footnote{IPN web page: http://www.ssl.berkeley.edu/ipn3/} 
  (\cite{IPN}), devoted to the
  localizing of GRBs by means of triangulation among different spacecrafts.

%______________________________________________________________

\section{Data processing and selection}

AGILE is on a nearly equatorial low-Earth orbit with a period 
of about 90 minutes; 
the MCAL data are sent once per orbit by the spacecraft 
to the ASI ground station based in Malindi,  Kenya, 
during the ground-station contact phase.
The data are processed as soon as they are received. 
The relevant data processing steps are energy evaluation,
data selection and photon list production.

\begin{itemize}

\item[-] {\em Gamma-ray energy release evaluation:} 
         The energy released in each bar is evaluated; 
         most of the events deposit energy in a single bar.  
         For multiple events, i.e., those involving more than one bar, 
         we attribute to the event the sum of the energies released 
         in all different bars.

\item[-] {\em On-ground data selection:} 
         the data stream transmitted to the ground 
         station for each GRB trigger is visually inspected, and spurious 
         triggers due to known electronic noise issues are discarded. Data in 
         which only very short timescale triggers (trigger on the 16-ms time 
         window or less) are activated are also carefully inspected but, since 
         they are mostly relevant for terrestrial gamma-ray flashes (TGF) science 
         (\cite{tgf}), they are not included in the present work.

\item[-] {\em Photon list production:} for each accepted trigger a photon list is 
         produced. The photon list contains the time of each detected photon, 
         and the energy released in each bar. 
         Many seconds of background data before and after the
         GRB data are included, as specified in the trigger logic configuration.

\end{itemize}
 
% --------------------------Long  Table 1 : general data:  
\addtocounter{table}{1}

This catalog includes 76 GRBs triggered on the 64-ms timescale and 
longer, plus nine GRBs triggered on the 16-ms timescale.
For the 64-ms sample 81\% of the GRBs activate the lower energy triggers 
$(E < 1.4 MeV)$ and only 17\% and 2\% the upper energy ones; the 
triggers for the four time windows of 64, 256, 1024 and 8192 ms represent 
12, 34, 29, and 25\% of the total. Triggers on timescales 
up to 16 ms where included only if the events were confirmed by at 
least another spacecraft belonging to the IPN. This latter criterion 
results in nine triggers on the 16 ms timescale only.

The GRBs included in the catalog are listed in Table \ref{maindata}.
The column \emph{contact} is the number of the AGILE orbit 
in which the GRB was detected. 
The \emph{trigger time} is the time of the GRB trigger signal 
(seconds from 1 January 2004); 
the celestial coordinates \emph{RA} and \emph{DEC} 
are given in decimal degrees and obtained from the relevant GCN circulars 
\footnote{GCN: A service of the  Astrophysics Science Division (ASD) at NASA's GSFC: http://gcn.gsfc.nasa.gov/}, 
the absence of decimals means that approximate values are obtained from the 
\emph{IPN} data available online (\cite{IPNFermi}).
The column  \emph{TH} is the angle between the AGILE pointing direction and the direction of the GRB. 
The column \emph{other detections} reports if the data for the GRB have been 
published also by other missions, or a GCN circular is issued for the detection.
The GRBs for which very high energy gamma-rays have been detected by the AGILE Tracker (TR) 
have been highlighted.

\begin{figure}
   \resizebox{\hsize}{!}{\includegraphics{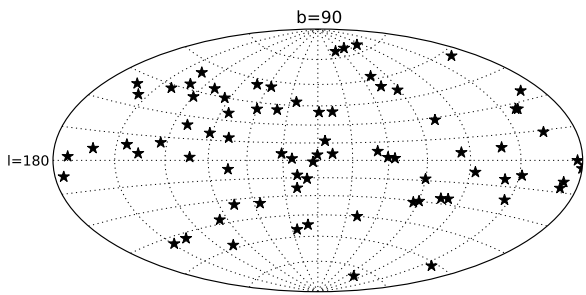}} 
   \caption{Sky distribution of MCAL GRBs in galactic coordinates. 
           The Galactic Center is at the center of the figure.}
   \label{figlb}
\end{figure}

 The MCAL is an all-sky instrument, but for geometrical reasons
 it is more sensitive to GRBs coming from the AGILE pointing direction.
 In the years 2008-2009 AGILE was used to follow gamma transients 
 pointing for many months on sources on the Galactic Plane and
 on Galactic Center. In Fig.~\ref{figlb}
 the GRB sky distribution is shown in galactic coordinates; a
 greater number of GRBs can be seen on the Galactic Center zone and
 around the Galactic Plane. 
 A list of AGILE pointings for most of the reference period is reported 
 by Pittori  (\cite{pointings}).
% See also  \ref{samaps} for a SuperAgile exposure map
% for about the same period 

%______________________________________________________________

\section{Temporal analysis}

For the temporal analysis of the GRBs we computed 
the $T_{90}$ and $T_{50}$ time intervals, 
defined as the intervals containing the 90\% and 50\%
of the total counts of gamma rays above the background. 
Our $T_{50}$ and $T_{90}$ measure the time width of the 
high energy part of the GRB (above $ \sim 300 \,keV$)
which is seen by MCAL.
We adopted the method of Koshut (\cite{koshut}):
\begin{itemize}
\item[-] the counts in the photon list are grouped in temporal bins 
         of 0.01 s; from which we obtained a light curve for the GRB. 

\item[-] By visual analysis of the light curve, 
         two background-only time intervals were identified, 
         one before and one after the burst. 
         We kept these time intervals as long as possible, to achieve
         a better background subtraction; 
         for their visual identification, 
         a binning of 0.1 sec was sometimes used, 
         which does not allow for an analysis of short GRBs, but gives less
         noisy time profiles. 

\item[-] The two selected intervals were used to define a background model
         in the GRB time interval by a low-order interpolating polynomial 
         (most of the time first order was sufficient).  
              
\item[-] The cumulative integral of the background-subtracted
         counts was computed for each time bin; 
         thus we obtained a cumulative time profile.    
         
\item[-] The mean values of the two background-only 
         intervals in the cumulative time profile were computed. 
         These values define a minimum and a maximum fluence level,
         expressed in counts.
         The mean value of the background region 
         before the GRB should theoretically be zero,
         but in practice, as also noticed by Koshut  (\cite{koshut}), 
         it can be different from zero, due to  background fluctuations. 
         
\item[-] A GRB fluence (in counts) is defined as the difference between the
         maximum and the minimum fluence level.
         The time interval between the points in which the cumulative time profile 
         is 5\% and 95\% of the GRB fluence defines the $T_{90}$ interval.
         In the same way, the time points at 25\% and 75\% define $T_{50}$.
         The uncertainties in $T_{50}$ and $T_{90}$ were computed following 
         Koshut (\cite{koshut}).
                            
\end{itemize} 

\begin{figure}
   \resizebox{\hsize}{!}{\includegraphics{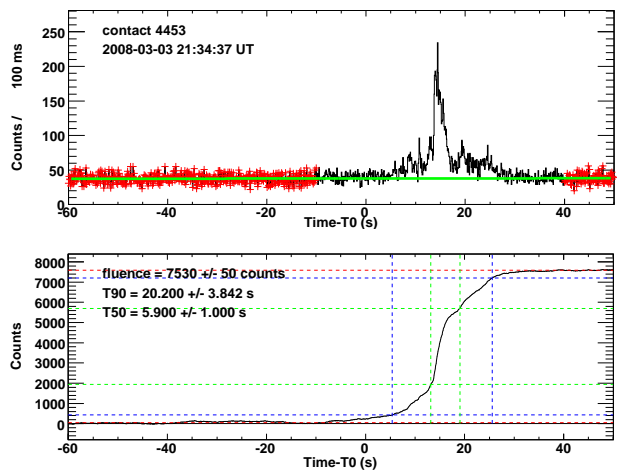}}
   \caption{GRB 080303B, time analysis.
           In the upper plot we show the GRB time profile; the two background zones
           are the time intervals: -60/-10 and 40/50 s.
           In the lower plot we show the cumulative 
           time profile; the lines defining the fluence, 
           $T_{50}$ and $T_{90}$ are shown as well.
           For a better visualization, a 0.1 s time binning has been
           used in this figure. }
   \label{fig5090}
\end{figure}

In Fig.~\ref{fig5090} an example of the procedure is shown;
in the upper panel we plot the time profile of  GRB 080803B along with the 
background zones and the line fitting the background values;
in the lower panel the cumulative integral is shown along with
the lines defining the fluence minimum and maximum levels, 
the $T_{90}$ and $T_{50}$ intervals.

% -------------------------- Long  Table 2 : timing data:  
\addtocounter{table}{1}

We performed a time analysis for 84 GRBs from our sample;
GRB 080603B (contact number 5750) is confirmed by the IPN network,
but too faint for a time analysis.
The results are shown in Table \ref{T5090}.  
$T_{50}$ and $T_{90}$, their starting times and the boundaries of the background intervals
are reported. All times are given in seconds measured from the trigger time. 
The column $T_{90}$ \emph{fluence} contains the  background-subtracted 
counts in the GRB region.
Our $T_{50}$ and $T_{90}$ measurements refer to the high-energy part of the GRB; 
for this reason the duration measured by other instruments, 
which are sensitive to lower energies, are often longer (see \cite{fermi2}):
we did not see the lower energy tails of some GRBs.

The main features of our sample 
are summarized in Table \ref{stats};
our average value for $T_{90}$ is about 13 sec. We had no
very long GRBs in our sample.

%-------------------------- Table 3 : timing summary 
\begin{table} 
\begin{center}
\begin{tabular}{llll}
\hline\hline 
   & $T_{90}$ (s) & $T_{50}$ (s) & Counts in $T_{90}$\\   
\hline 
 min. value & 0.04 &  0.02 &    66 \\ 
 max. value &89.81 & 21.44 & 12472 \\ 
 average    &12.61 &  4.64 &  1822 \\   
\hline
\end{tabular}
\caption{GRB temporal analysis summary}
\label{stats}
\end{center}
\end{table}

\begin{figure}
   \resizebox{\hsize}{!}{\includegraphics{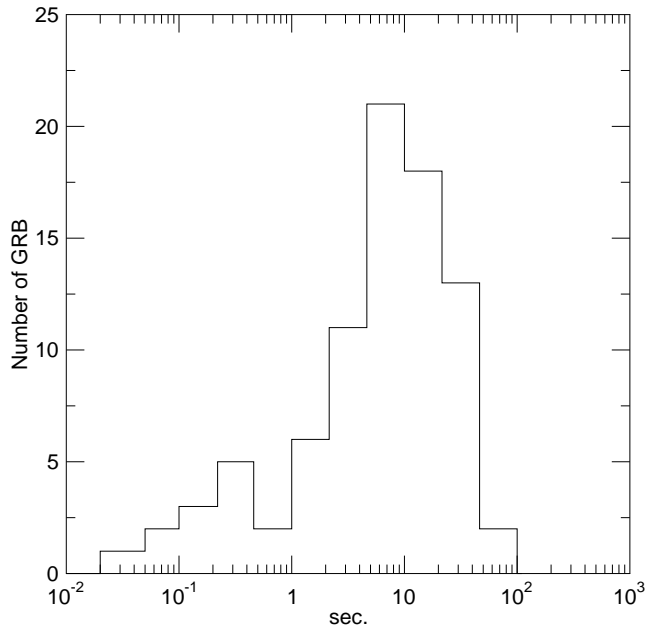}}
   \caption{$T_{90}$ distribution.}
   \label{figT90}
\end{figure}

\begin{figure}
   \resizebox{\hsize}{!}{\includegraphics{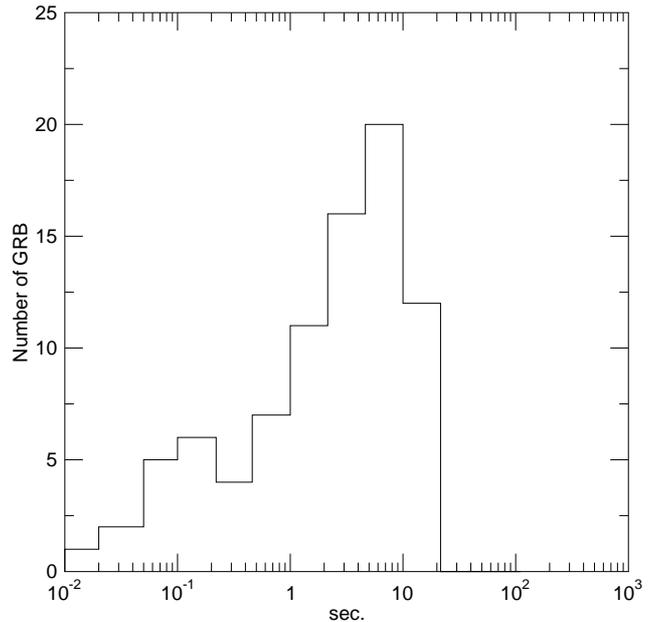}} 
   \caption{$T_{50}$ distribution.}
   \label{figT50}
\end{figure}

The distributions of $T_{90}$ and $T_{50}$ are shown 
in Fig.~\ref{figT90} and Fig.~\ref{figT50}.
The data are consistent with the well-known bimodal distribution (\cite{bimodal}). 
We see a clear peak for long GRBs and a peak for short GRBs 
that is less resolved due to the limited statistics for this group. 
The two maxima are observed at
$T_{90}\sim 0.3 $ s and $T_{90}\sim 8 $ s. 
Considering as \emph{short} 
the GRB with $T_{90} < 2 $ s,
we have 21\% of short GRB, a value consistent with 
the results of instruments operating in the hard X-ray range: Paciesas 
reported a value of 18\% for the Fermi data (with an uncertainty of 3-4\%)  
and of 24\% for the BATSE catalog (\cite{fermi2});
see also Kouveliotou (\cite{bimodal}) and Frontera (\cite{SAX}) for previous samples.

%______________________________________________________________

\section{Spectral analysis}

Only a subsample of the GRBs in the catalog have enough photons to allow 
for a spectral analysis. Background subtraction is a critical 
point in the analysis, and we can obtain reliable spectra only for
GRBs with more than $\sim 1000$ background-subtracted counts in 
the $T_{90}$ interval.
The MCAL effective area has a cutoff below
 $\sim 350$ keV, for this reason a fit of our data with a
Band function (\cite{band}) is, in most cases, not suitable, since we cannot constrain
the low-energy part of the spectra. We instead used a simple power law
function to describe the GRB high-energy part between 400 keV
and 5 MeV: $ f(E)=N\times (E/E0)^{-\beta} $,
with a photon index $\beta $ that is similar to the $\beta $
index in the GRB Band model. 
We processed here all  GRBs in the same way, fitting with the same 
simple power-law function, to obtain a set of homogeneous results.
A detailed analysis for some high
flux GRBs, such as GRB 090510, has been published elsewhere 
(\cite{grb080514B}, \cite{grb090510}).

The uncertainty in the response matrix for gamma-rays coming from
below the instrument is higher. These gamma-rays cross the service module of the
satellite before hitting MCAL.
For this reason we considered only the GRBs with a 
direction of less than 90 degrees
from the AGILE pointing direction. 

Our procedure consisted of the following steps:
\begin{itemize}

\item[-] {\em Obtaining the spectra of the detected GRB photons 
         from the photon list}.
         The photon list produced by the MCAL onboard software
         was used to obtain the spectrum of energy released in MCAL 
         by the GRB during the $T_{90}$ time interval.
         A background spectrum was also extracted from the data,
         summing over a time interval as long as possible before the trigger 
         (30--50 s). A large background interval smoothes the short-time oscillations 
         in the background, which can introduce errors when the burst fluence is low.  
         The energies released in all bars that correspond to a single event, 
         were summed; the bars with an energy lower than 400 keV 
         were discarded in this phase 
         to minimize the contribution of possible uncertainties
         in diode calibrations, which mostly affect the low-energy events.          
         The spectra were binned at a fixed bin size
         of 50 keV, up to 10 MeV (200 bins).
         The spectra were produced in the FITS format (\cite{fits}).
                  
\item[-] {\em Dead-time correction}. The anti-coincidence (AC) 
         veto system of the AGILE tracker (\cite{AC})
         is active during the GRB.
         This system also affects the MCAL counts, 
         blocking MCAL for $ 5.4 \, \mu$s for each AC event. 
         With a particle background rate of $\sim 10$ kHz
         and a GRB photon rate of $\sim 15-30$ kHz,
         this dead-time effect amounts to about $ 10-20 \% $.
         To account for this effect,
         we considered an average of the AC rates 
         during the GRB and the background interval 
         and decreased the exposure time in a consistent way.  
            
\item[-] {\em Fitting}.
         { The XSPEC program (\cite{xspec}) was used for fitting. 
             We used the XSPEC CSTAT statistic, a modified
             version of the Cash statistic (\cite{cstat}), 
             which is more appropriate when counts are low. 
         A response matrix and
         auxiliary response files were computed for each GRB
         with the same energy binning 
         of the spectra and considering the angle
         between the GRB direction and the AGILE axis. 
         The discarding of bars with a low energy deposit was accounted for.
         The method used for the computation of the response matrix
         will be detailed in a forthcoming article (\cite{calibrazione}).
         \par\noindent
         A power law function $f(E)=N \times (E/500 \, keV)^{-\beta}$ 
         was used for the fits over the energy range 400--5000 keV.         
         }

\end{itemize} 

\begin{figure}
   \resizebox{\hsize}{!}{\includegraphics{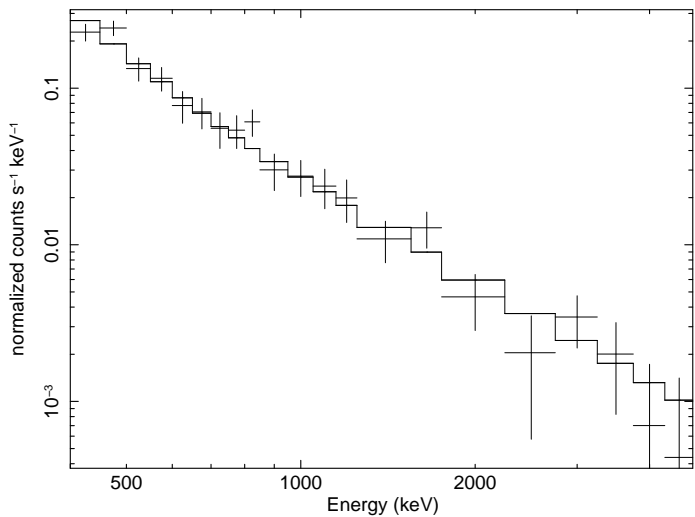}}
   \caption{GRB 080916C: spectral data and folded model from XSPEC
            (continuous line). The power law model is $ \sim E^{-2.24}$.}  
            
   \label{figpower} 
\end{figure}

In Fig.~\ref{figpower}, an example of a GRB background-subtracted
spectrum is shown along with the folded model 
from the XSPEC fit (continuous line); 
the error bars are one-sigma poisson errors. 
Some energy bins over  $\sim 1.5$ MeV were merged. 

% -------------------------- Long  Table 4 : spectral data:  
\addtocounter{table}{1} 

Spectral data are shown in Table \ref{spectra}, where 
the spectral index $\beta$ and the normalization parameter $N$ are 
reported along with
the model flux integrated over the energy interval $400-5000$ keV
and averaged over $T_{90}$. 
We also report the fluence, integrated over $T_{90}$,
the  $CSTAT$ value, and the number of degrees of freedom (dof)
from the XSPEC fit.  

\begin{figure}
   \resizebox{\hsize}{!}{\includegraphics{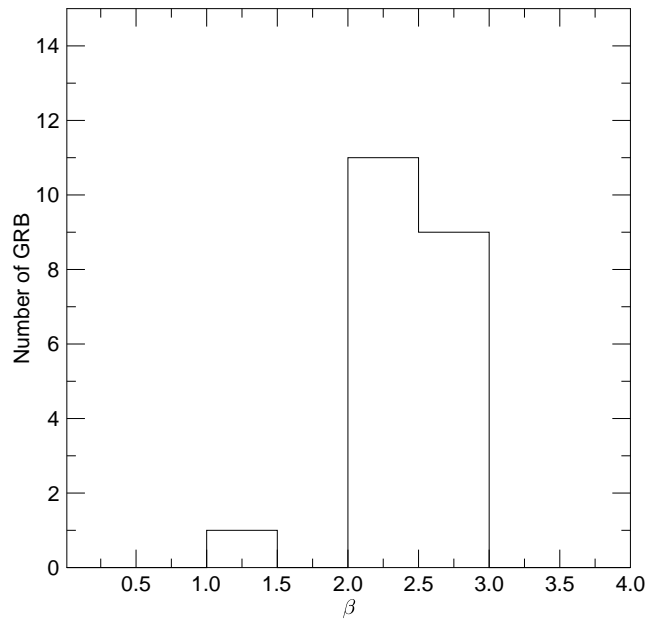}} 
   \caption{Power law index $\beta$ distribution.}
   \label{beta} 
\end{figure}

\begin{figure}
   \resizebox{\hsize}{!}{\includegraphics{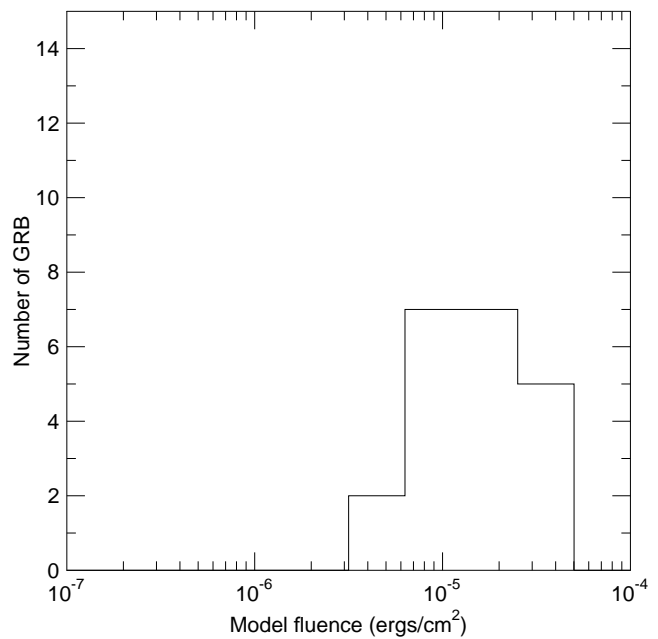}}
   \caption{Fluence distribution.}
   \label{flux} 
\end{figure}

In Fig.~\ref{beta} we show the distribution
of the power law index $\beta$. 
Most of the values are between 2 and 3.
A similar distribution is found for the Band model $\beta$  parameter 
in the {\it Fermi} sample of Bissaldi (\cite{fermibiss}) or
in the {\it Fermi} spectral catalog (\cite{fermispectra}),
but we see a somewhat greater number of high $\beta$ GRBs. 
We are still investigating this effect; 
a detailed comparison of our spectral data with other samples 
will be presented in a forthcoming paper (\cite{calibrazione}).
The model fluence distribution is shown in Fig.~\ref{flux}.

%______________________________________________________________

\section{Conclusions}

We presented the GRB data collected with the AGILE MCAL instrument
between February 2008 and  October 2009. 
The effective area of MCAL decreases below $\sim 350$ keV,
and this GRB sample consists of hard events, with data up to the MeV range.   
Our temporal analysis shows a $T_{90}$ distribution consistent with
other studies, in spite of our higher energy range.
We also presented a spectral analysis for a subsample of our data:
we fitted our data with a simple power law. 
Our spectral index is comparable to the $\beta$
parameter in the classical Band GRB model. Our results are similar to other
studies, but with  somewhat lower spectral indexes.  

\begin{acknowledgements}
AGILE is a mission of the Italian Space Agency (ASI), with co-participation
of INAF (Istituto Nazionale di Astrofisica) and INFN (Istituto
Nazionale di Fisica Nucleare). This work was carried out in the frame
of the ASI-INAF agreement I/028/12/0.

Alessandro Ursi participated to a first step of the data analysis.
\end{acknowledgements} 

\bibliographystyle{aa}

%%%% Long Table 1 : MAIN GRB DATA  

\clearpage \onecolumn   
{\small
\longtab{1}{
\begin{longtable}{rclccccl}
\caption{ \textbf{Main GRB data}. 
          Boldface GRBs have high-energy photons detected by the AGILE Tracker.
          For the detection by other missions the 
          following acronyms are used: \emph{INT} for the INTEGRAL satellite (\cite{INTEGRAL}),
   \emph{RH} for RHESSI (\cite{RHESSI}), 
   \emph{SW} for the SWIFT satellite (\cite{SWIFT}),
   \emph{KW} for KONUS/Wind  (\cite{KONUS}),
   \emph{SU} for Suzaku (\cite{SUZAKU}),
   \emph{FE} for {\it Fermi} (\cite{fermilat}, \cite{fermigbm}),
   \emph{SA} for SuperAGILE, 
   \emph{TR} for the AGILE Tracker,
   \emph{OPT} means optical or radio followup, 
   \emph{IPN} means that a tentative GRB position was obtained by  
        \emph{IPN} data  or that a GCN was issued by the \emph{IPN}.} \\
\hline\hline
 Contact &  Trigger time & Name & Date  & RA   & DEC  & TH   & Other detections \\
 \multicolumn{1}{c}{ID}  & sec.from 01-01-2004 &      & UTC   & deg. & deg. & deg. &                 \\
\hline
\\
\endfirsthead
\caption{continued. \textbf{Main GRB data.}} \\
\hline\hline
 Contact                &  Trigger time & Name & Date  & RA   & DEC  & TH   & Other detections \\
 \multicolumn{1}{c}{ID} & sec.from 01-01-2004 &      & UTC   & deg. & deg. & deg. &                 \\
\hline
\\
\endhead
\hline
\endfoot
4172 &  129942289.420  &    080212B      &  2008-02-12 23:04:49.420  &     123     &    22      & 98 &     IPN \\
4453 &  131664877.825  &    080303B      &  2008-03-03 21:34:37.825  &     270     &     -26     & 108 &    IPN \\
4673 &  133014356.656  &    080319C      &  2008-03-19 12:25:56.656  &     259.0     &     55.4     & 84 &    INT KW SU SW OPT \\
4798 &  133776179.856  &    080328       &  2008-03-28 08:02:59.856  &     80.5     &     47.5     & 155 &    INT KW SU SW OPT \\
4946 &  134685725.108  &    080407       &  2008-04-07 20:42:05.108  &     183     &     20     & 76 &    KW  IPN\\
5030 &  135199576.492  &    080413       &  2008-04-13 19:26:16.492  &     287.3     &     -27.7     & 49 &    KW SW OPT\\
5370 &  137281445.197  &    080507B      &  2008-05-07 21:44:05.197  &     247     &     -24     & 95 &    IPN \\
{\bf 5462} & 137843756.822 & {\bf 080514B } & 2008-05-14 09:55:56.822  &     322.8     &     0.7     & 37 &    KW SU IPN/SA TR SW OPT\\
5665 &  139084070.650  &    080528       &  2008-05-28 18:27:50.650  &     77     &     28     & 93 &     KW IPN \\
5692 &  139248696.743  &    080530       &  2008-05-30 16:11:36.743  &     186     &    0     & 69 &     IPN\\
5750 &  139606694.696  &      080603B       &  2008-06-03 19:38:14.696  &     176.5     &     68.1     & 74 &    INT KW SW OPT \\
5799 &  139903643.150  &    080607       &  2008-06-07 06:07:23.150  &     194.9     &     15.9     & 114 &    INT KW RH SW OPT \\
5837 &  140136578.898  &    080609       &  2008-06-09 22:49:38.898  &     262     &     -26     & 132 &     IPN \\
5852 &  140231946.977  &    080611       &  2008-06-11 01:19:06.977  &     83     &     27     & 85 &    IPN \\
5887 &  140440358.003  &    080613B      &  2008-06-13 11:12:38.003  &     173.8     &     -7.1     & 39 &    INT KW SW OPT \\
6044 &  141402060.830  &    080624       &  2008-06-24 14:21:00.830  &     ...     &     ...     & ... &      \\
6258 &  142712152.973  &    080709       &  2008-07-09 18:15:52.973  &     245     &     -22    & 91 &    IPN \\
6328 &  143142769.240  &    080714       &  2008-07-14 17:52:49.240  &     188.1     &     -60.3     & 13 &    KW SW OPT \\
6345 &  143246921.106  &    080715       &  2008-07-15 22:48:41.106  &     214.7     &     65.9     & 66 &    IPN FE\\
6422 &  143720706.723  &    080721       &  2008-07-21 10:25:06.723  &     224.5     &     -11.7     & 49 &    KW RH INT SW \\
6452 &  143904198.133  &    080723B      &  2008-07-23 13:23:18.133  &     176.8     &     -60.2     & 13 &    INT KW SA SW OPT\\
6458 &  143941060.261  &    080723D      &  2008-07-23 23:37:40.261  &    105.3       &   71.1       & 134 &    FE  \\
6506 &  144231210.935  &    080727B      &  2008-07-27 08:13:30.935  &     276.9     &     1.2     & 101 &    KW SW OPT\\
6600 &  144807750.714  &    080803B      &  2008-08-03 00:22:30.714  &     108     &     22     & 50 &    IPN \\
6800 &  146029932.250  &    080817       &  2008-08-17 03:52:12.250  &     151.3     &     -19.3     & 57 &    FE \\
6919 &  146758428.560  &    080825C      &  2008-08-25 14:13:48.560  &     232.2     &     -4.9     & 75 &    FE \\
7073 &  147700735.012  &    080905       &  2008-09-05 11:58:55.012      &     287.67    &    -18.87    & 139 &   FR SW OPT \\
7083 &  147762311.241  &    080906B      &  2008-09-06 05:05:11.241  &     182.8     &     -6.4     & 115 &    FE  \\
7221 &  148608766.336  &    080916C      &  2008-09-16 00:12:46.336  &     119.8     &     -56.6     & 86 &    FE IPN KW SW OPT \\
7255 &  148815876.398  &    080918       &  2008-09-18 09:44:36.398  &     22     &     14     & 107 &    IPN \\
7445 &  149980665.476  &    081001       &  2008-10-01 21:17:45.476  &     276.6     &     -8.7     & 21 &    KW SA SW OPT \\
7483 &  150211576.844  &    081004       &  2008-10-04 13:26:16.844  &     113     &     -3     & 147 &    IPN \\
7889 &  152700294.277  &    081102B      &  2008-11-02 08:44:54.277  &     231.2     &    35.2     & 53 &    FE \\
7935 &  152976706.752  &    081105       &  2008-11-05 13:31:46.752  &     4.0     &     3.5     & 66 &    IPN SW \\
8006 &  153411942.992  &    081110       &  2008-11-10 14:25:42.992  &     123.6     &     21.2     & 124 &    FE \\ 
8165 &  154384526.221  &    081121       &  2008-11-21 20:35:26.221  &     89.3     &     -60.6     & 141 &     FE KW SW OPT \\
8216 &  154698820.603  &    081125       &  2008-11-25 11:53:40.603  &     42.7     &     -18.9     & 84 &    FE \\
8330 &  155397093.361  &    081203B      &  2008-12-03 13:51:33.361  &     228.8     &     44.4     & 69 &    KW SU SW OPT  \\
8420 &  155950316.320  &    081209       &  2008-12-09 23:31:56.320  &     45.3     &     63.5     & 52 &    FE IPN KW  \\
8502 &  156451717.870  &    081215       &  2008-12-15 18:48:37.870  &     135.0     &     53.8     & 89 &    FE IPN KW \\
8593 &  157006438.184  &    081222       &  2008-12-22 04:53:58.184  &     22.75    &    -34.10    & 88 &    FE KW SW OPT \\
8631 &  157238275.145  &    081224       &  2008-12-24 21:17:55.145  &     201.7     &     75.1     & 59 &    FE SU SW \\
8712 &  157735152.060  &    081230B      &  2008-12-30 15:19:12.060  &     ...     &     ...     & ... &    \\
8747 &  157949728.040  &    090102       &  2009-01-02 02:55:28.040  &     128.2     &     33.1     & 94 &    KW SW \\
9156 &  160452584.144  &    090131       &  2009-01-31 02:09:44.144  &     353.0     &     16.4     & 55 &    FE SU  \\ 
9179 &  160595218.112  &    090201       &  2009-02-01 17:46:58.112  &     92.0     &     -46.6     & 139 &     KW SW OPT \\
9260 &  161088448.472  &    090207B      &  2009-02-07 10:47:28.472  &     ...     &     ...     & ... &    \\
9508 &  162605543.649  &    090225       &  2009-02-25 00:12:23.649  &     358.1     &     61.0     & 19 &   FE \\
9553 &  162881600.901  &    090228       &  2009-02-28 04:53:20.901  &     106.8     &     -24.3     & 111 &    FE \\
9621 &  163300476.010  &    090305B      &  2009-03-05 01:14:36.010  &     155.1     &     68.1     & 120 &     FE\\
9731 &  163969232.676  &    090312       &  2009-03-12 19:00:32.676  &     ...     &     ...     & ... &    \\
9951 &  165317810.892  &    090328       &  2009-03-28 09:36:50.892  &     90.9     &     -41.9     & 105 &    FE KW SW \\
9955 &  165344824.903  &    090328B      &  2009-03-28 17:07:04.903  &     155.7     &     33.4     & 132 &    FE IPN \\
9972 &  165446602.621  &    090329       &  2009-03-29 21:23:22.621  &     ...       &     ...      & ... &    \\
{\bf 10007} & 165659724.000  & {\bf 090401B } & 2009-04-01 08:35:24.000  &     95.09     &     -8.96     & 41 &    KW SA SU TR SW OPT\\
10139 &  166467473.377  &   090410       &  2009-04-10 16:57:53.377  &     334.96     &     15.41     & 54 &    SU SW OPT\\
10335 &  167667129.448  &   090424       &  2009-04-24 14:12:09.448  &     189.5     &     16.8     & 103 &    FE SU SW OPT \\
10364 &  167848413.139  &   090426C      &  2009-04-26 16:33:33.202  &      82.7     &     -9.7     & 143 &  FE \\
10383 &  167959586.511  &   090427       &  2009-04-27 23:26:26.511  &     235.9     &     13.5     & 71 &    IPN KW \\
{\bf 10553} & 168999780.445  & {\bf 090510 } & 2009-05-10 00:23:00.445  &     333.56 &     -26.61   & 61 &    FE SU SW TR OPT \\
10741 &  170155676.304  &   090523       &  2009-05-23 09:27:56.304  &     30        &     11       & 108 &   IPN  \\
10814 &  170598152.774  &   090528B      &  2009-05-28 12:22:32.774  &     312.2     &     32.7     & 28 &     FE SU \\
10832 &  170707316.602  &   090529C      &  2009-05-29 18:41:56.602  &     ...       &     ...      & ... &    \\
10860 &  170879756.431  &   090531B      &  2009-05-31 18:35:56.492  &    252.07     &   -36.02     & 82 &   FE SW OPT \\ 
10999 &  171732798.963  &   090610       &  2009-06-10 15:33:25.936  &     84.2      &     35.4     & 84 &  FE \\ 
11027 &  171903051.103  &   090612       &  2009-06-12 14:50:51.103  &     81.1     &     17.8      & 92 &     FE \\
11092 &  172299598.562  &   090617       &  2009-06-17 04:59:58.562  &     78.9     &     15.7      & 91 &    FE SU \\
11108 &  172398504.772  &   090618       &  2009-06-18 08:28:24.772  &     294.0     &     78.4     & 38 &    FE KW SA SU SW OPT\\
11137 &  172575386.596  &   090620       &  2009-06-20 09:36:26.596  &     237.4     &     61.2     & 57 &    FE \\
11404 &  174209913.464  &   090709       &  2009-07-09 07:38:33.464  &     289.9     &     60.7     & 48 &    KW SA SU SW OPT \\
11513 &  174876569.696  &   090717       &  2009-07-17 00:49:29.696  &     86.8     &     -64.2     & 152 &    FE SU \\
11514 &  174883231.619  &   090717B      &  2009-07-17 02:40:31.619  &     247.0     &     23.0     & 68 &    FE \\
11541 &  175051886.412  &   090719       &  2009-07-19 01:31:26.412  &     341.3     &     -67.9     & 133 &    FE KW \\
11565 &  175194176.902  &   090720B      &  2009-07-20 17:02:56.905   &   203.0    &  -54.8  & 155 &  FE KW IPN \\
11851 &  176945288.503  &   090809B      &  2009-08-09 23:28:08.503  &     93.5     &     0.1     & 89 &    FE \\
11914 &  177324581.259  &   090814C      &  2009-08-14 08:49:41.259  &     332.5     &     58.9     & 154 &    FE \\
11993 &  177813526.664  &   090820       &  2009-08-20 00:38:46.664  &     87.7     &     27.0     & 130 &    FE OPT\\
12044 &  178128673.156  &   090823       &  2009-08-23 16:11:13.156  &     128.67     &     60.6     & 141 &    IPN KW SW \\
12107 &  178511020.971  &   090828       &  2009-08-28 02:23:40.971  &     124.4     &     -26.1     & 72 &    FE \\
12129 &  178646894.468  &   090829       &  2009-08-29 16:08:14.468  &     329.2     &     -34.2     & 64 &    FE SU \\
12156 &  178811964.864  &   090831D      &  2009-08-31 13:59:24.864  &     ...       &     ...       & ... &    \\
12269 &  179505087.102  &   090908       &  2009-09-08 14:31:27.102  &     ...       &     ...       & ... &    \\
12466 &  180709001.651  &   090922       &  2009-09-22 12:56:42.137   &  17.1    &  74.3   & 117 &  FE \\
12616 &  181629345.746  &   091003       &  2009-10-03 04:35:45.746  &     251.1     &     37.2     & 67 &    FE KW SU SW OPT \\
12714 &  182227390.920  &   091010       &  2009-10-10 02:43:10.920  &     298.7     &     -22.5     & 9 &    FE KW SA SU SW OPT \\
\hline 
\label{maindata}
\end{longtable}  
}}
%
%%%% Long Table 2 : temporal data 
%
\clearpage \onecolumn 
{\small
\longtab{2}{
\begin{longtable}{rlcccclcccc}
\caption{GRB temporal data. } \\
\hline\hline
 Contact                  & Name & $T_{90}$  & $T_{90}$ start   & $T_{50}$ & $T_{50}$ start     & $T_{90}$ Fluence & \multicolumn{4}{c}{Background intervals}  \\
 \multicolumn{1}{c}{ID}   &      & sec.    & sec. from trigger  & sec.     & sec. from trigger  &  counts & \multicolumn{4}{c}{sec. from trigger} \\
\hline
\\
\endfirsthead
\caption{continued. GRB temporal data.}\\
\hline\hline
 Contact                 & Name & $T_{90}$  & $T_{90}$ start       & $T_{50}$ & $T_{50}$ start       & $T_{90}$ Fluence & \multicolumn{4}{c}{Background intervals} \\
 \multicolumn{1}{c}{ID}  &      & sec.    & sec. from trigger  & sec.   & sec. from trigger  &  counts & \multicolumn{4}{c}{sec. from trigger}  \\
\hline
\\
\endhead
\hline
\endfoot
4172 & 080212B &    6.16 $\pm$  3.83  &   0.265 &   2.11  $\pm$  0.46  &   1.125 &  1931 $\pm$ 154 &   -60   &  -10   &   20   & 30 \\ 
4453 & 080303B &   20.09 $\pm$  3.45  &   5.345 &   5.84  $\pm$  0.96  &  13.165 &  6775 $\pm$ 215 &   -30   &  -10   &   40   & 50 \\ 
4673 & 080319C &   19.22 $\pm$  3.64  & -10.815 &   4.41  $\pm$  3.36  &   0.295 &  1101 $\pm$ 142 &   -30   &  -15   &   15   & 30 \\ 
4798 & 080328 &    15.42 $\pm$ 11.14  &   4.355 &  11.83  $\pm$  2.53  &   6.425 &  1079 $\pm$ 102 &   -30   &   -5   &   25   & 40 \\ 
4946 & 080407 &    20.16 $\pm$  1.86  &   2.215 &   8.72  $\pm$  0.74  &   6.905 &  7427 $\pm$ 159 &   -30   &  -15   &   30   & 50 \\ 
5030 & 080413 &    30.28 $\pm$ 15.59  &  -8.395 &   7.13  $\pm$  2.47  &   7.175 &   873 $\pm$ 144 &   -30   &  -10   &   25   & 30 \\ 
5370 & 080507B &    0.22 $\pm$  0.25  &   0.045 &   0.03  $\pm$  0.04  &   0.225 &    83 $\pm$  18 &    -0.5 &    0   &    1   & 2 \\ 
{\bf 5462} & 080514B &    6.09 $\pm$  4.74  &   0.245 &   3.73  $\pm$  0.96  &   2.075 &  1347 $\pm$ 117 &   -20   &   -5   &   10   & 20 \\ 
5665 & 080528 &     0.24 $\pm$  1.71  &   0.025 &   0.11  $\pm$  0.06  &   0.065 &   296 $\pm$  48 &   -10   &   -1   &    1   & 7 \\ 
5692 & 080530 &     1.43 $\pm$  2.19  &  -0.365 &   0.16  $\pm$  0.14  &   0.215 &   457 $\pm$  51 &    -5   &   -1   &    1   & 5 \\ 
5799 & 080607 &    10.38 $\pm$  3.03  &   1.175 &   4.75  $\pm$  0.49  &   4.585 &  4165 $\pm$ 119 &   -80   &  -10   &   20   & 30 \\ 
5837 & 080609 &    20.60 $\pm$  0.61  &   1.105 &  10.42  $\pm$  0.92  &   6.115 & 12472 $\pm$ 180 &   -80   &  -10   &   30   & 45 \\ 
5852 & 080611 &     0.04 $\pm$  0.06  &   0.035 &   0.02  $\pm$  0.01  &   0.055 &   342 $\pm$  24 &   -50   &   -1   &    1   & 20 \\ 
5887 & 080613B &   18.33 $\pm$  6.71  &  -0.045 &   9.24  $\pm$  2.65  &   1.645 &  1465 $\pm$ 147 &   -20   &   -5   &   20   & 40 \\ 
6044 & 080624 &    23.64 $\pm$  8.28  & -14.565 &  12.52  $\pm$  3.18  & -10.165 &  1005 $\pm$ 149 &   -40   &  -20   &   20   & 30 \\ 
6258 & 080709 &    20.94 $\pm$  1.84  &   1.825 &  11.74  $\pm$  0.50  &   4.225 &  7944 $\pm$ 163 &   -40   &  -10   &   30   & 50 \\ 
6328 & 080714 &    39.65 $\pm$ 20.64  &  -7.085 &  11.99  $\pm$  8.12  &   3.565 &  1021 $\pm$ 158 &   -18   &  -10   &   35   & 50 \\ 
6345 & 080715 &    26.39 $\pm$  7.72  &  -6.065 &   7.06  $\pm$ 17.82  &  -0.505 &   464 $\pm$ 137 &   -25   &  -10   &   20   & 40 \\ 
6422 & 080721 &    15.64 $\pm$  1.13  &   1.045 &   5.97  $\pm$  0.51  &   9.355 &  5460 $\pm$ 198 &   -25   &  -10   &   20   & 40 \\ 
6452 & 080723B &   45.65 $\pm$  1.09  & -43.975 &  13.44  $\pm$  1.03  & -14.365 &  3443 $\pm$ 206 &   -70   &  -50   &   10   & 40 \\ 
6458 & 080723D &   43.56 $\pm$  2.70  &   4.115 &   8.40  $\pm$  5.83  &  12.445 &  3930 $\pm$ 235 &   -70   &  -10   &   60   & 70 \\ 
6506 & 080727B &    1.99 $\pm$  1.18  &  -0.135 &   0.45  $\pm$  1.12  &   0.095 &   227 $\pm$  54 &    -5   &   -1   &    5   & 10 \\ 
6600 & 080803B &   17.50 $\pm$ 13.55  &   0.285 &   5.47  $\pm$  4.47  &   3.895 &   617 $\pm$ 144 &   -40   &   -5   &   20   & 30 \\ 
6800 & 080817 &    17.84 $\pm$  2.03  &  -0.625 &   9.44  $\pm$  0.77  &   3.795 &  4239 $\pm$ 175 &   -30   &  -10   &   30   & 60 \\ 
6919 & 080825C &    3.67 $\pm$  1.29  &   0.605 &   1.51  $\pm$  0.25  &   1.425 &  1069 $\pm$  69 &   -10   &   -1   &    7   & 15 \\ 
7073 & 080905 &     1.26 $\pm$  3.53  &   0.005 &   0.80  $\pm$  0.79  &   0.135 &   164 $\pm$  42 &   -10   &   -1   &    2   &  6 \\
7083 & 080906B &    4.78 $\pm$  6.87  &   0.555 &   0.95  $\pm$  3.50  &   0.865 &   317 $\pm$  94 &   -20   &   -3   &   10   & 20 \\ 
7221 & 080916C &   39.31 $\pm$  7.35  &   0.695 &  21.44  $\pm$  5.16  &   3.985 &  3778 $\pm$ 215 &   -60   &  -10   &   50   & 80 \\ 
7255 & 080918 &     4.64 $\pm$  2.87  &  -0.915 &   0.85  $\pm$  0.15  &   1.285 &  1395 $\pm$  89 &    -8   &   -3   &    5   & 10 \\ 
7445 & 081001 &     5.40 $\pm$  3.10  &  -1.495 &   1.24  $\pm$  1.75  &   0.165 &   500 $\pm$  68 &    -6   &   -2   &    7   & 10 \\ 
7483 & 081004 &     0.13 $\pm$  1.73  &   0.005 &   0.04  $\pm$  0.05  &   0.015 &   197 $\pm$  45 &    -5   &   -1   &    1   & 5 \\ 
7889 & 081102B &    7.30 $\pm$ 14.06  &   3.525 &   1.70  $\pm$  2.08  &   5.945 &   346 $\pm$  74 &   -30   &    0   &   15   & 50 \\ 
7935 & 081105 &     6.17 $\pm$  1.63  &   0.045 &   2.50  $\pm$  0.68  &   1.265 &  3154 $\pm$ 107 &   -30   &   -2   &    8   & 40 \\ 
8006 & 081110 &     3.64 $\pm$  1.76  &  -2.000 &   0.64  $\pm$  0.38  &   0.135 &   489 $\pm$  44 &   -30   &   -2   &   25   &  40 \\
8165 & 081121 &    18.96 $\pm$  2.61  &   0.935 &   6.38  $\pm$  2.40  &   7.915 &  1805 $\pm$ 123 &   -25   &   -1   &   25   & 40 \\ 
8216 & 081125 &     4.12 $\pm$  5.34  &  -2.225 &   1.94  $\pm$  1.27  &  -1.135 &   360 $\pm$  60 &   -25   &   -3   &    5   & 25 \\ 
8330 & 081203B &    20.16 $\pm$ 14.88  &   0.955 &   5.28  $\pm$  2.60  &   6.885 &  1027 $\pm$ 116 &   -40   &   -4   &   20   & 40 \\ 
8420 & 081209 &     0.13 $\pm$  0.29  &   0.045 &   0.06  $\pm$  0.04  &   0.055 &   184 $\pm$  30 &    -0.5 &   -0.2 &    0.4 & 0.8 \\ 
8502 & 081215 &     4.92 $\pm$  0.98  &  -0.585 &   2.46  $\pm$  0.40  &   0.385 &  2365 $\pm$ 118 &   -20   &   -5   &   10   & 20 \\ 
8593 & 081222 &     6.69 $\pm$  6.71  &   0.995 &   3.48  $\pm$  2.95  &   2.125 &   362 $\pm$  85 &   -5    &   -1   &   10   & 15 \\ 
8631 & 081224 &     7.80 $\pm$  2.81  &   0.385 &   2.45  $\pm$  0.46  &   1.135 &  2706 $\pm$ 133 &   -20   &   -5   &   15   & 30 \\ 
8712 & 081230C &    2.52 $\pm$  1.81  &   0.045 &   1.19  $\pm$  1.39  &   0.395 &   225 $\pm$  60 &   -5    &   -1   &    5   & 10 \\ 
8747 & 090102 &    14.19 $\pm$  3.00  &   5.205 &   9.27  $\pm$  1.43  &   8.455 &  1137 $\pm$ 106 &   -20   &   -5   &   25   & 40 \\ 
9156 & 090131 &    32.41 $\pm$ 11.12  & -16.790 &   9.74  $\pm$ 19.45  &   0.100 &   624 $\pm$ 154 &   -28   &  -20   &   15   & 40 \\ 
9179 & 090201 &    25.47 $\pm$  8.81  & -12.045 &  10.83  $\pm$  2.43  &   1.105 &  2200 $\pm$ 195 &   -40   &  -20   &   20   & 40 \\ 
9260 & 090207B &    0.10 $\pm$  0.07  &  -0.005 &   0.06  $\pm$  0.04  &   0.005 &    66 $\pm$  13 &   -1    &   -0.2 &    0.2 & 0.3 \\ 
9508 & 090225 &     0.28 $\pm$  0.16  &  -0.005 &   0.09  $\pm$  0.13  &   0.095 &    82 $\pm$  15 &   -0.5  &    0   &    0.4 & 0.8 \\ 
9553 & 090228 &     0.09 $\pm$  0.53  &   0.005 &   0.06  $\pm$  0.02  &   0.015 &   316 $\pm$  29 &   -0.8  &   -0.2 &    0.4 & 0.8 \\ 
9621 & 090305B &    0.53 $\pm$  0.97  &  -0.145 &   0.27  $\pm$  0.13  &   0.025 &   186 $\pm$  33 &  -10    &   -2   &    1   & 5 \\ 
9731 & 090312 &     8.49 $\pm$ 14.91  &   2.545 &   4.45  $\pm$  2.04  &   5.325 &   636 $\pm$  77 &  -22    &   -5   &   15   & 35 \\ 
9951 & 090328 &    26.98 $\pm$  4.59  &   0.275 &  10.72  $\pm$  1.78  &   6.605 &  4420 $\pm$ 175 &  -25    &   -5   &   35   & 50 \\ 
9955 & 090328B &    0.15 $\pm$  0.22  &  -0.025 &   0.07  $\pm$  0.04  &   0.005 &   167 $\pm$  26 &  -0.6   &   -0.2 &    0.3 & 1 \\ 
9972 & 090329 &     4.15 $\pm$  4.13  &  -1.505 &   1.39  $\pm$  0.40  &   0.205 &   834 $\pm$  37 &  -18    &   -4   &   10   & 30 \\
{\bf 10007} & 090401B &   8.98 $\pm$  0.29  &   0.915 &   1.6   $\pm$  0.19  &   7.355 &  4041 $\pm$ 119 &  -10    &   -1   &   12   & 22 \\ 
10139 & 090410 &   12.19 $\pm$ 12.60  &   2.035 &   4.07  $\pm$  4.61  &   5.065 &   592 $\pm$ 102 &   -5    &    0   &   20   & 30 \\ 
10335 & 090424 &    6.78 $\pm$  6.41  &  -2.755 &   2.96  $\pm$  0.47  &   0.355 &   854 $\pm$  92 &  -20    &   -5   &    5   & 15 \\ 
10364 & 090426C &   4.17 $\pm$  4.60  &  -1.395 &   1.17  $\pm$  1.56  &   0.425 &   216 $\pm$  63 &   -8    &   -2   &   15   & 25 \\
10383 & 090427 &    2.97 $\pm$  1.03  &   8.045 &   0.98  $\pm$  0.06  &   8.825 &  6325 $\pm$ 179 &  -20    &   -5   &   15   & 30 \\ 
{\bf 10553} & 090510 &    5.19 $\pm$  5.91  &   0.065 &   0.29  $\pm$  0.22  &   0.125 &  1553 $\pm$  64 &  -20    &   -5   &   15   & 30 \\ 
10741 & 090523 &    1.12 $\pm$  1.05  &  -0.855 &   0.13  $\pm$  0.12  &   0.055 &   188 $\pm$  33 &  -25    &   -1   &    2   & 10 \\ 
10814 & 090528B &  30.36 $\pm$ 19.88  &   1.285 &  15.92  $\pm$  9.16  &   7.205 &  1271 $\pm$ 151 &  -30    &   -5   &   35   & 50 \\ 
10832 & 090529D &  15.83 $\pm$  8.09  &  -5.005 &   3.73  $\pm$ 11.31  &   4.355 &   386 $\pm$  81 &  -30    &   -5   &   15   & 32 \\ 
10860   &  090531B  &    1.66  $\pm$  2.06 & -0.975 &    0.43 $\pm$  0.38 & 0.015 &  258 $\pm$  54 &  -50    &   -2   &   10   & 40 \\ 
10999 & 090610 &   8.82  $\pm$  7.38  &  -1.415 & 8.82    $\pm$  7.38  &   3.165 &   293 $\pm$  77 &  -25    &   -2   &   20   & 45 \\
11027 & 090612 &   3.12  $\pm$  8.38  &  -0.845 &   1.32  $\pm$  0.81  &   0.125 &   716 $\pm$ 109 &  -25    &   -5   &    5   & 25 \\ 
11092 & 090617 &   0.67  $\pm$  1.17  &  -0.305 &   0.20  $\pm$  0.31  &  -0.005 &    83 $\pm$  19 &   -4    &   -0.3 &    0.3 & 4 \\ 
11108 & 090618 &  89.81  $\pm$ 11.74  &   3.055 &   9.54  $\pm$  3.20  &  63.285 & 10614 $\pm$ 271 &  -30    &   -5   &  140   & 150 \\ 
11137 & 090620 &  13.61  $\pm$  3.76  &  -7.695 &   3.67  $\pm$  2.55  &  -1.275 &   737 $\pm$ 101 &  -40    &  -10   &   15   & 25 \\ 
11404 & 090709 &  53.22  $\pm$ 11.00  &   3.385 &  20.67  $\pm$  2.88  &  23.465 &  4242 $\pm$ 225 &  -40    &  -10   &   70   & 90 \\ 
11513 & 090717 &   8.42  $\pm$ 12.02  &   3.375 &   5.19  $\pm$  3.68  &   4.875 &   460 $\pm$  74 &  -15    &   -5   &   15   & 30 \\ 
11514 & 090717B &  0.37  $\pm$  1.06  &  -0.035 &   0.11  $\pm$  0.11  &   0.135 &    87 $\pm$  20 &   -3    &   -1   &    1   & 2 \\ 
11541 & 090719 &   7.28  $\pm$ 13.24  &   0.275 &   3.97  $\pm$  1.03  &   1.715 &  2860 $\pm$ 181 &  -20    &  -15   &   20   & 40 \\ 
11565 & 090720B &  6.21  $\pm$  0.14  &   0.015 &   5.73  $\pm$  0.33  &   0.135 &  1234 $\pm$  83 &  -30    &   -9   &   25   & 45 \\
11851 & 090809B &  6.81  $\pm$ 14.56  &   6.655 &   1.71  $\pm$  1.04  &   8.055 &   877 $\pm$ 114 &  -10    &   -1   &   20   & 30 \\ 
11914 & 090814C &  0.28  $\pm$  1.64  &  -0.075 &   0.11  $\pm$  0.11  &  -0.005 &   137 $\pm$  28 &  -10    &   -0.5 &    1   & 2 \\ 
11993 & 090820 &   8.18  $\pm$  2.03  &   1.055 &   3.36  $\pm$  0.25  &   2.435 &  6454 $\pm$ 123 &  -30    &   -10  &   20   & 40 \\ 
12044 & 090823 &   5.33  $\pm$  8.37  &  -0.325 &   2.44  $\pm$  1.26  &   0.925 &   806 $\pm$ 107 &  -20    &   -5   &   10   & 25 \\ 
12107 & 090828 &   8.46  $\pm$  9.72  &  -3.175 &   2.90  $\pm$  1.72  &   0.125 &   537 $\pm$  88 &  -20    &   -5   &   10   & 25 \\ 
12129 & 090829 &  23.24  $\pm$  5.19  &   1.595 &   7.04  $\pm$  1.14  &   7.895 &  4244 $\pm$ 174 &  -30    &   -5   &   35   & 50 \\ 
12156 & 090831D &  2.10  $\pm$  0.23  &  -2.005 &   0.73  $\pm$  1.10  &  -0.695 &   169 $\pm$  15 &   -6    &   -2   &   15   & 20 \\
12269 & 090908C &  4.05  $\pm$  6.57  &   0.065 &   0.72  $\pm$  0.68  &   0.305 &   380 $\pm$  54 &  -10    &   -5   &    5   & 10 \\ 
12466 &  090922 & 15.92  $\pm$  5.24  &  -9.045 &   4.71  $\pm$ 10.00  &  -0.075 &   466 $\pm$  87 &  -60    &  -10   &   20   & 30 \\
12616 & 091003  & 24.68  $\pm$ 11.40  &  -4.125 &  16.45  $\pm$  1.17  &   2.365 &  2397 $\pm$ 170 &  -40    &  -10   &   30   & 45 \\ 
12714 & 091010 &   3.17  $\pm$  1.95  &  -1.495 &   2.24  $\pm$  1.93  &  -1.225 &   260 $\pm$  67 &   -8    &   -3   &    4   & 8 \\ 
\hline 
\label{T5090}
\end{longtable}  
}}

% 
%%%% 
%
%%%% Long Table 4 : spectral data  
%
\clearpage \onecolumn 
{\small
\longtab{4}{
\begin{longtable}{rlcccccr}
\caption{GRB spectral data. } \\
\hline\hline
 Contact          & Name  & $\beta$ &         $N$         & Flux             & Fluence      & $CSTAT$  & dof \\
  \multicolumn{1}{c}{ID} &       &         & $ph./cm^2/s/keV$  & $ergs/cm^2/s$  & $ergs/cm^2$  &           &     \\
         &       &         &  at 500 keV         & 400-5000 keV     & in $T_{90}$  &           &     \\
\hline
\\
\endfirsthead
\caption{continued. GRB spectral data.}\\
\hline\hline
 Contact                 & Name  & $\beta$ &         $N$         & Flux             & Fluence      & $CSTAT$  & dof   \\
 \multicolumn{1}{c}{ID}  &       &         & $ph./cm^2/s/keV$  & $ergs/cm^2/s$  & $ergs/cm^2$  &           &       \\
         &       &         &  at 500 keV         & 400-5000 keV     & in $T_{90}$  &           &       \\
\hline
\\  
\endhead 
\hline
\endfoot 
 4673 & 080319C & 2.26 $+$ 0.43 - 0.34 & 5.42E-04 $\pm$ 1.13E-04 & 4.25e-07 & 8.17e-06 & 85.28 & 89 \\ 
 4946 & 080407  & 2.72 $+$ 0.08 - 0.08 & 3.02E-03 $\pm$ 8.15E-05 & 1.65e-06 & 3.33e-05 & 116.66 & 89 \\ 
{\bf 5462}\footnotemark[1]   & 080514B & 2.61 $+$ 0.25 - 0.21 & 1.64E-03 $\pm$ 5.14E-05 & 9.68e-07 & 5.90e-06 & 84.31 & 89 \\ 
 5887 & 080613B & 2.19 $+$ 0.22 - 0.20 & 5.45E-04 $\pm$ 6.15E-05 & 4.57e-07 & 8.38e-06 & 113.42 & 89 \\ 
 6328 & 080714  & 2.14 $+$ 1.24 - 0.83 & 2.09E-04 $\pm$ 7.27E-05 & 1.83e-07 & 7.26e-06 & 59.23 & 89 \\ 
 6452 & 080723B & 2.93 $+$ 0.34 - 0.29 & 6.26E-04 $\pm$ 9.22E-05 & 3.00e-07 & 1.37e-05 & 90.09 & 89 \\ 
 6800 & 080817  & 2.47 $+$ 0.11 - 0.11 & 1.52E-03 $\pm$ 6.64E-05 & 1.00e-06 & 1.78e-05 & 107.25 & 89 \\ 
 6919 & 080825C & 2.39 $+$ 0.17 - 0.19 & 2.36E-03 $\pm$ 9.37E-05 & 1.65e-06 & 6.06e-06 & 69.03 & 89 \\ 
 7221 & 080916C & 2.24 $+$ 0.15 - 0.14 & 1.33E-03 $\pm$ 1.32E-04 & 1.06e-06 & 4.17e-05 & 90.58 & 89 \\ 
 7935 & 081105  & 2.21 $+$ 0.09 - 0.09 & 2.94E-03 $\pm$ 1.80E-04 & 2.42e-06 & 1.49e-05 & 138.05 & 89 \\ 
 8330 & 081203B  & 2.65 $+$ 0.36 - 0.30 & 6.08E-04 $\pm$ 6.97E-05 & 3.50e-07 & 7.06e-06 & 100.44 & 89 \\ 
 8502 & 081215  & 2.43 $+$ 0.10 - 0.10 & 1.04E-02 $\pm$ 4.15E-04 & 7.07e-06 & 3.48e-05 & 86.40 & 89 \\ 
 8631 & 081224  & 2.79 $+$ 0.12 - 0.11 & 3.28E-03 $\pm$ 1.66E-04 & 1.71e-06 & 1.33e-05 & 97.05 & 89 \\ 
{\bf 10007}\footnotemark[2] & 090401B & 2.14 $+$ 0.07 - 0.07 & 2.79E-03 $\pm$ 1.48E-04 & 2.45e-06 & 2.20e-05 & 84.37 & 89 \\ 
10383 & 090427  & 2.30 $+$ 0.05 - 0.05 & 1.55E-02 $\pm$ 3.48E-04 & 1.18e-05 & 3.50e-05 & 108.55 & 89 \\ 
{\bf 10553}\footnotemark[3] & 090510  & 1.43 $+$ 0.11 - 0.11 & 1.08E-03 $\pm$ 7.28E-04 & 2.16e-06 & 1.12e-05 & 71.59 & 89 \\ 
10814 & 090528B & 2.55 $+$ 0.31 - 0.27 & 4.45E-04 $\pm$ 6.90E-05 & 2.75e-07 & 8.35e-06 & 75.48 & 89 \\ 
11108 & 090618  & 2.73 $+$ 0.27 - 0.32 & 6.32E-04 $\pm$ 6.33E-05 & 3.43e-07 & 3.08e-05 & 78.96 & 89 \\ 
11404 & 090709  & 2.66 $+$ 0.19 - 0.17 & 6.51E-04 $\pm$ 6.36E-05 & 3.71e-07 & 1.97e-05 & 96.29 & 89 \\ 
12129 & 090829  & 2.38 $+$ 0.12 - 0.11 & 1.14E-03 $\pm$ 5.35E-05 & 8.07e-07 & 1.88e-05 & 130.16 & 89 \\ 
12616 & 091003  & 2.72 $+$ 0.26 - 0.23 & 7.57E-04 $\pm$ 7.11E-05 & 4.13e-07 & 1.02e-05 & 109.16 & 89 \\ 
\hline
\\
\multicolumn{7}{l}{\footnotemark[1]Detailed analysis published (\cite{grb080514B}).} \\
\multicolumn{7}{l}{\footnotemark[2]Detailed analysis published (\cite{grb090401B}, \cite{grb090401Bbis}).}\\
\multicolumn{7}{l}{\footnotemark[3]Detailed analysis published (\cite{grb090510}).} \\
\\
\hline 
\label{spectra}
\end{longtable}  
}} 

%
%%%%

\end{document}